# Crystalline symmetry-dependent magnon formation in itinerant ferromagnet SrRuO$_3$

H. I. Seo[*], S. Woo[*], S. G. Jeong, Prof. W. S. Choi

Department of Physics, Sungkyunkwan University, Suwon 16419, Republic of Korea

e-mail : choiws@skku.edu

J. Kim[*], Prof. T. Park

Department of Physics, Sungkyunkwan University, Suwon 16419, Republic of Korea

Center for Quantum Materials and Superconductivity, Department of Physics, Sungkyunkwan University, Suwon 16419, Republic of Korea

[*]These authors contributed equally to this paper.

**ABSTRACT**

SrRuO$_3$ (SRO) is an itinerant ferromagnet with strong coupling between the charge, spin, and lattice degrees of freedom. This strong coupling suggests that the electronic and magnetic behaviors of SRO are highly susceptible to changes in the lattice distortion. Here we show how the spin interaction and resultant magnon formation change with the modification in the crystallographic orientation. We fabricated SRO epitaxial thin films with (100), (110), and (111) surface orientations, to systematically modulate the spin interaction and spin dimensionality. The reduced spin dimensionality and enhanced exchange interaction in the (111)-oriented SRO thin film significantly suppresses magnon formation. Our study comprehensively demonstrates the facile tunability of magnon formation and spin interaction in correlated oxide thin films.

## I. INTRODUCTION

Perovskite SrRuO$_3$ (SRO) is an itinerant ferromagnet (critical temperature, $T_c$ ~160 K) with intriguing electronic and magnetic characteristics that are strongly coupled [1,2]. Below $T^*$ ~20 K, SRO shows Fermi liquid (FL) behavior in which Coulomb interaction serves as the main scattering mechanism for itinerant electrons [3]. At $T^* < T < T_c$, however, magnon excitations are known to induce magnon-electron interactions, leading to non-Fermi liquid (NFL) behavior [4]. More recently, neutron scattering provided the formation of magnons in SRO [5,6], supporting the magnon-electron interactions in the NFL phase.

In SRO, the charge-spin coupling can be further tuned by modifying the lattice degree of freedom, including the crystalline symmetry. This makes the material attractive for tailoring various functional properties [7,8]. An unusual Hall effect [9], a dimensional crossover of the electromagnetic ground states [10], and an enhancement in electrocatalytic activity [11] have been observed in SRO epitaxial thin films with modulated lattice structures. Specifically, modifications of the epitaxial strain and crystalline surface symmetry in SRO have been employed for facile control over its lattice distortions, and in turn, its electronic structures and exchange interactions. As a prominent example, $T_c$ was found to increase from 150 to 154 to 156 K as the surface orientation of the SRO thin films on SrTiO$_3$ (STO) substrates changed from (100) to (110) to (111) [12-14].

Indeed, the enhancement of $T_c$ can be understood in terms of the distinctive lattice distortions induced by changes in the surface orientations [12,14-17]. In general, the use of identical thin film and substrate materials leads to the same degree of epitaxial strain resulting from the lattice mismatch. However, even in such cases, the structural symmetry of the thin film can change owing to the modified surface symmetry. In particular, cubic substrates such as STO reveal square, rectangular, and triangular surface crystalline symmetries when exposed in the (100), (110), and (111) surface orientations, respectively (Figs. 1(a)-1(c)) [13]. Such changes in the surface crystalline symmetry in SRO epitaxial thin films necessarily affect the microscopic Ru-O bonding geometry and orbital overlap, leading to substantial changes in the spin ordering and dynamics [18-20].

In this study, we show that magnon formation can be systematically modulated using crystalline symmetry engineering in SRO epitaxial thin films. We controlled the crystalline symmetry of SRO by employing the epitaxial strain with distinct surface orientations. In this way, we were able to tune the spin dimensionality and spin interaction, which we extracted from the magnetization measurement result using the scaling law and Bloch's law [21]. The lattice distortion-dependent spin dimensionality and interaction led to significant changes in the temperature range of the magnon-dominated region, which was identified through both transport and magnetization measurements for different orientations and external magnetic fields [15]. The findings of the study can be summarized as follows: (1) Confirmation of the enhancement of $T_c$ for the (110) and (111)-oriented SRO thin films compared to the (100)-oriented one [12,14-17]. (2) Increase of the spin dimensionality and (3) exchange interaction as the orientation changes from (100) to (110) and (111) [16,21,22]. (4) Introduction of $T_m$, a characteristic temperature of magnon, and its decreasing behavior as the orientation changes from (100) to (110) and (111). (5) Systematic increase of $T^*$ as the orientation changes from (100) to (110) and (111) [22].

## II. EXPERIMENTS

Epitaxial single-crystalline SRO thin films were grown on atomically-flat single-crystal STO substrates with (100), (110), and (111) surface orientations using pulsed laser epitaxy at 750 °C and 100 mTorr of oxygen partial pressure. Before deposition, the surfaces of the STO substrates were treated using buffered HF and then annealed at 1000 °C for 6 h [23]. Stoichiometric ceramic SRO was used as a target. An excimer (KrF) laser (248 nm; IPEX 868, Lightmachinery) with a fluence of 1.5 J/cm$^2$ and repetition rate of 5 Hz was used for ablation. Three samples with (100), (110), and (111) orientations were grown at the same time to minimize the stoichiometry and defect deviations between the samples [24]. The lattice structures of the SRO thin films were characterized by high-resolution X-ray diffraction (XRD, X'Pert PRO, Malvern Panalytical). The thickness of the thin films (~30 nm) was measured by X-ray reflectometry (XRR). The temperature- ($T$-) dependent resistivity, $\rho$ ($T$), was measured from 300 to 1.8 K by employing a physical property measurement system (PPMS, PPMS-9T, Quantum Design), using the four-probe method with Pt electrodes and Au wires. The $T$-dependent magnetization, $M$ ($T$), was measured from 300 to 1.8 K at the magnetic fields of 100 Oe and 5 T along the out-of-plane direction of the thin films (field-cooled warming), using a magnetic property measurement system (MPMS, MPMS-XL, Quantum Design).

**III. RESULTS AND DISCUSSION**

The lattice distortions of SRO were modulated using epitaxially strained thin film geometries with different surface orientations. Figs. 1(a)-(c) show the schematic lattice structures of the SRO thin films grown on (100)-, (110)- and (111)-oriented STO substrates [15,25]. As the surface symmetries change, the lattice distortions of the SRO epitaxial thin films change as well [26,27]. The (100)-, (110)- and (111)-oriented SRO thin films undergo tetragonal, monoclinic, and trigonal distortions, respectively [17]. The X-ray diffraction (XRD) $\theta$–$2\theta$ scans (Fig. 1(d), also see Supplemental Material Fig. S1) [24] and reciprocal space maps (Fig. 1(e)) confirm the high-quality growth of the epitaxially strained SRO thin films [12,18,25]. Fig. 1(f) summarizes the atomic interplanar spacing, $d_{inter}$, obtained from the XRD $\theta$–$2\theta$ results. $d_{inter}$ naturally decreases as the surface orientation changes from (100) to (110) to (111). Unexpectedly, however, the perovskite pseudocubic unit cell volume, $V_{pc}$, also decreases systematically from 60.2 to 60.1 to 59.9 Å$^3$ in (100)-, (110)-, and (111)-oriented SRO thin films, respectively. These changes in $V_{pc}$ manifests the distinctive lattice distortions in SRO epitaxial thin films with different surface orientations [16,21].

Disparate lattice distortions in the SRO thin films lead to changes in the tendency of the spin alignment. Figs. 2(a), (b), and (c) show the normalized magnetization, $M(T)/M(2\,\text{K})$, for the (100)-, (110)-, and (111)-oriented SRO thin films, respectively. Open circles denote the experimental data under 100 Oe magnetic fields applied along the out-of-plane direction. The magnetic field along the out-of-plane direction was chosen, because it was consistently reported as the magnetic easy axis of the SRO thin films on STO substrate, regardless of the surface orientations [13,14,21,28,29]. While a slight tilt away from the perpendicular axis or small difference in the actual magnitude of magnetic anisotropy between the SRO thin films with different orientations might be possible, these would not affect the magnetic ordering of the thin films significantly, as the magnetic easy axis dominantly determines the magnetic anisotropy. The green solid and black dashed lines are the fits obtained using the scaling and the Bloch's laws, respectively. The sudden upturns of $M(T)$ at 147, 155, and 157 K in the (100)-, (110)- and (111)-oriented SRO thin films, respectively, mark the $T_c$, which was determined as the dip in the $(1/M)*(dM/dT)$ of $M(T)$ (see Supplemental Material Fig. S3) [24]. The scaling law (green solid lines), $M(T) \sim A(T_c - T)^\gamma$, approximates a ferromagnetic phase transition at $T_c$ and provides a good fit for the critical region just below $T_c$ [30,31]. Here, $A$ is a parameter related to the magnetic susceptibility and $\gamma$ is the critical exponent associated with the spin dimensionality. The $\gamma$ value of an Ising- (Heisenberg-) type ferromagnet is theoretically given as $0.326 \pm 0.004$ ($0.36 \pm 0.03$) [31], corresponding to the spin dimension of 1D (3D). In general, $\gamma$ increases as the dimensionality of the spins increases. The scaling law fits of $M(T)$ give $\gamma = 0.311$, $0.287$, and $0.259$, for (100)-, (110)-, and (111)-oriented SRO, respectively, implying a systematic reduction of spin dimensionality in SRO with the change in lattice distortion and decrease in $V_{pc}$.

The fitting obtained using Bloch's law provided a negative correlation between the magnon excitation and spin exchange coupling constant ($J$). Bloch's law, $M(T) \sim 1 - BT^{1.5}$, implies a reduction in $M$ due to the formation of thermally excited magnons, and provides a good fit around $T = 0$ [32]. Bloch's law allows us to extract $J$ between two neighboring Ru atoms through $B = (0.0587/S)(k_B/2JS)^{3/2}$, where $S$ is the total spin of Ru$^{4+}$ ($S = 1$) [21,30,33]. We note that the change in the spin dimensionality could modify the conventional Bloch's law. The effect was estimated by introducing a magnon gap ($\Delta$) into Bloch's law (see Appendix in Supplemental Material), and the small deviation was reflected as an error (~10%) in obtaining the $J$ value [24]. The (100)-oriented SRO thin film has the lowest $J$ value of 15.8 $k_B$ K, which systematically increases to 24.0 and 26.9 $k_B$ K in the (110)- and (111)-oriented SRO thin films, respectively [21,22]. We further defined $T_m$ (Figs. 2(a)-(c)) as the $T$ at which the deviation between the values from the experimental $M(T)$ and the Bloch's law fit exceeds the standard deviation of the experimental $M(T)$ values. $T_m$ decreases from 124 to 80 to 75 K as the surface orientation of the thin film changes from (100) to (110) to (111), respectively. The higher $T_m$ in the (100)-oriented SRO thin film compared to the other orientations indicates that the magnon contribution to $M$ persists up to a higher $T$ [30,32]. From $J$ values, we estimated the spin wave stiffness constant ($D = (1/3)SJza^2$), where $z$ and $a$ are the number of nearest neighbors and the nearest neighboring distance, respectively [5,6,34]. $S = 1$ was assumed for the SRO thin films. The $D$ systematically increases from 321.2, 487.9, and 546.9 meVÅ$^2$ as the surface orientation changes from (100) to (110) to (111), respectively. Fig. 2(d) summarizes the fitting parameters obtained from the $M(T)$ results and their fittings, exhibiting the systematic trends described.

The distinctive spin interactions in the SRO thin films also affect the electron scattering, picked up by the transport measurements (Fig. 2(e)). $T_c$ is characterized by an anomaly in $\rho(T)$ at ~150 K due to the reduced spin-disorder scattering below $T_c$ [12,16,22]. The $T_c$ of (100)-, (110)-, and (111)-oriented SRO thin films estimated from $\rho(T)$ are 144, 154, and 155 K [12,23], respectively, which show a consistent trend with the values obtained from $M$ ($T$) (Fig. 2(f)) [14]. The RRR values, i.e., $\rho_{300\,K} / \rho_{2\,K}$ were 6.2, 11.1, and 11.1 of (100)-, (110)-, and (111)-oriented SRO thin films, respectively, indicating good quality of the SRO thin film in Fig. S2 [13,14]. To further identify the effect of spin interaction on the electronic transport, we plotted $d\rho / dT$ as a function of $T$ (Fig. 3(a)). As previously discussed, the electronic transport in the FL phase is dominated by the electron-electron scattering, which leads to the critical exponent of $n = 2.0$ where $n = d\ln(\rho\,(T) - \rho_0) / d\ln T$ ($\rho_0$ is the residual resistivity) [35]. On the contrary, the NFL phase in the SRO is dominated by the electron-magnon scattering, which leads to the critical exponent of $n = 1.5$ [16,36]. From $\rho$ ($T$), we defined $T^*$ (see the arrows in Fig 3(a)) as the $T$ corresponding to $n = 1.75$, which marks the broad transition between the NFL and FL phases [4]. $T^*$ increases systematically from 20.4 to 21.9 to 23.9 K as the surface orientation changes from (100) to (110) to (111), respectively, as shown in Fig. 3(b). The enhanced $T^*$ in the (111)-oriented SRO corresponds to a stronger spin interaction at low $T$.

We summarize the extracted parameters and characteristic $T$s in Table I. From the characteristic $T$s, we further plot the values of $n$ as a function of orientation and $T$ as shown in Fig. 3(b). It is evident that the low spin dimensionality and strong spin interaction suppress the magnon formation in the (111)-oriented SRO thin film, which has the lowest $T_m$ and the highest $T^*$. In contrast, the (100)-oriented SRO thin film is more susceptible to magnon excitation and has the largest $T$ window. Overall, the diagram in Fig. 3(b) consistently shows the correlation between the lattice distortion, spin ordering, and magnon formation in SRO thin films.

To further validate the above findings, we obtained the characteristic $T$s under an external magnetic field of $H\perp = 5$ T [24], which sufficiently saturates all the magnetic moments in the SRO thin films [13,14,21,28,29], as shown in Figs. 3(c) and 3(d). In general, high external magnetic fields tend to align the spins and suppress spin-wave formation. This trend is exhibited by the results under $H\perp = 5$ T, in which $T_m$ ($T^*$) is decreased (increased) for all the SRO thin films (see Supplemental Material Fig. S4). In addition to the drastically reduced $T$ window of the magnon excitation, the orientation dependence is more or less washed out owing to the strong tendency for spin alignment at 5 T.

## IV. CONCLUSION

In summary, we investigated the surface orientation-dependent magnon formation in SRO

epitaxial thin films. By changing the surface orientation of the STO substrate, we achieved tetragonal, monoclinic, and trigonal symmetry in (100)-, (110)-, and (111)-oriented SRO thin films, respectively. As the orientation changed from (100) to (110) to (111), a decrease in the spin dimensionality and an increase in the spin exchange interaction were observed. These led to a drastic change in the $T$ window for magnon formation, i.e., the spin-wave excitation was significantly suppressed in the (111)-oriented SRO thin film. Our study provides a means for the facile control of spin interaction and magnon formation in SRO thin films and heterostructures, leading to a better understanding of the fundamental electronic and magnetic properties of this intriguing system.

**ACKNOWLEDGMENTS**

This work was supported by the Basic Science Research Programs through the National Research Foundation of Korea (NRF) (NRF-2019R1A2B5B02004546 and NRF-2012R1A3A2048816 (J. H. K., and T. P.)). We thank Jung Hoon Han for his critical advice and discussion about the estimation of spin dimensionality and exchange coupling.

**FIGURES & TABLES**

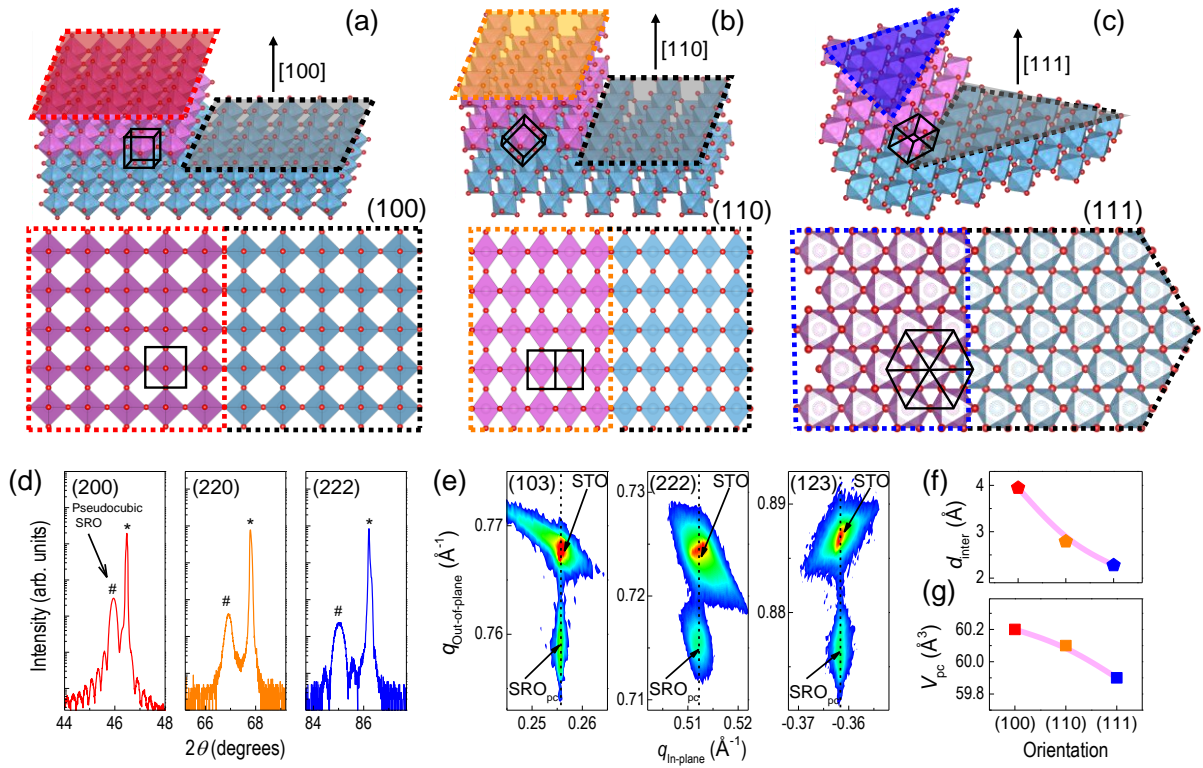

**FIG. 1.** Lattice structures of (100)-, (110)-, and (111)-oriented SRO epitaxial thin films. (a)–(c) Schematic diagrams of the SRO thin films with distinct surface orientations. The bottom panel shows the top view, and hence, represents the in-plane geometry of the SRO films and substrates. The black square, rectangles, and triangles indicate the surface symmetry of each orientation. (d) XRD $\theta$–$2\theta$ scans of the SRO thin films on (100)-, (110)-, and (111)-oriented STO substrates, respectively. * and # indicate STO substrates and SRO thin films, respectively. (e) The reciprocal space maps of the (100)-, (110)-, and (111)-oriented SRO thin films, shown around the (103), (222), and (123) Bragg reflections of the STO substrate, respectively. (f) Interplanar distance ($d_{\text{inter}}$) and perovskite pseudocubic unit cell volume ($V_{\text{pc}}$) of the SRO thin films. The lines are guides to the eye and the sizes of the symbols represent the error bars.

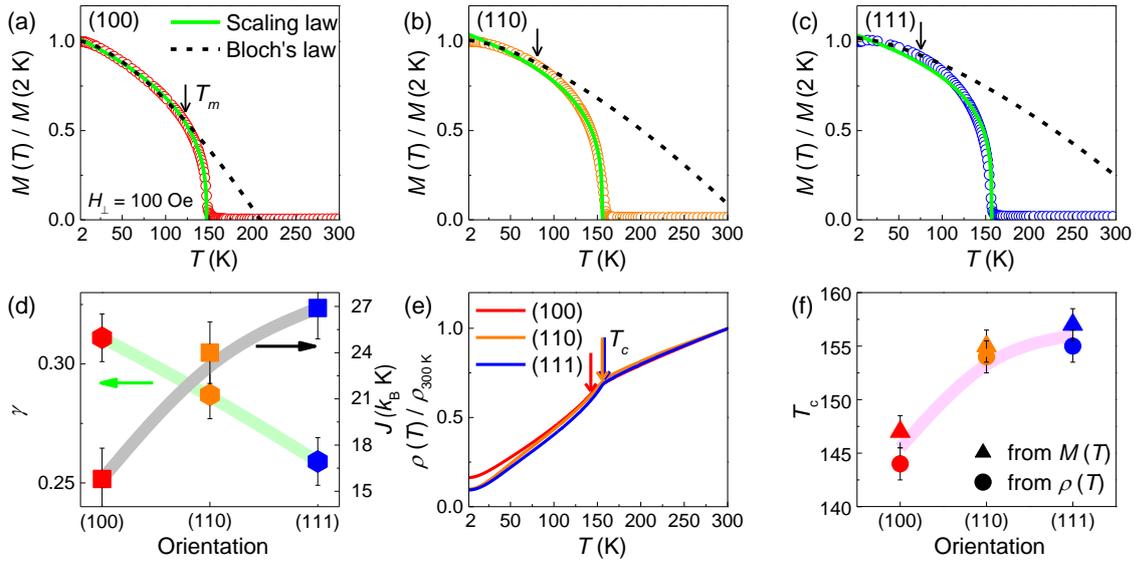

**FIG. 2.** $M(T)/M(2\,\text{K})$ of (a) (100)-, (b) (110)-, and (c) (111)-oriented SRO thin films. A 100 Oe magnetic field was applied along the out-of-plane direction. Green solid (black dashed) lines are the fits using the scaling law (Bloch' Law). The arrows indicate $T_m$. (d) Critical exponent $\gamma$ and exchange coupling constant $J$ of the SRO thin films. The lines are guides to the eye. (e) $\rho(T)/\rho_{300\,\text{K}}$ for the (100)-, (110)-, and (111)-oriented SRO films. The arrows indicate $T_c$. (f) The triangles and circles represent $T_c$ obtained from $M(T)$ and $\rho(T)$, respectively.

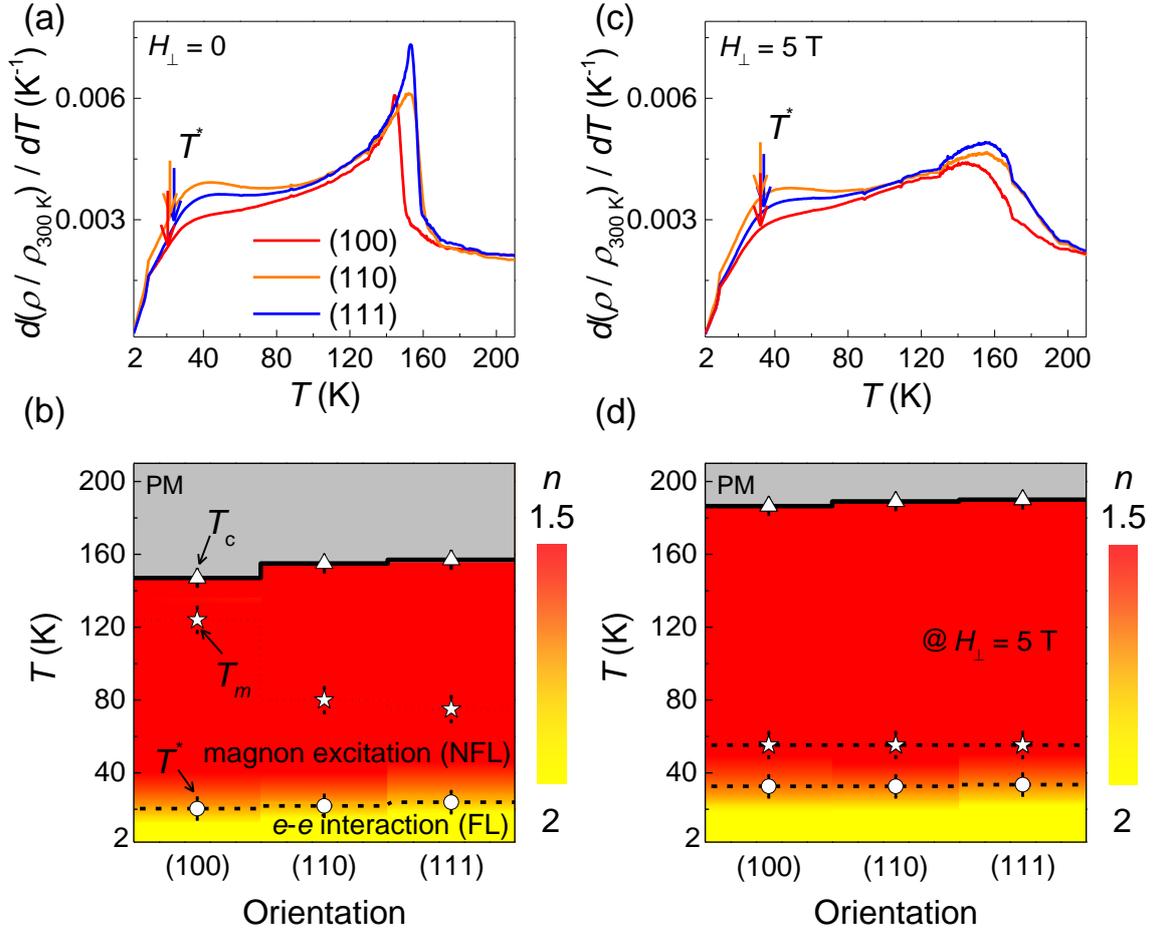

**FIG. 3.** (a) $T$ derivative of $\rho$ ($T$) for the SRO thin films. The arrows indicate the phase transition temperature $T^*$ between the NFL and FL phases. (b) Critical exponents $n$ as functions of $T$ are shown for the (100)-, (110)-, and (111)-oriented SRO thin films. The triangles, stars, and circles indicate $T_c$, $T_m$, and $T^*$, respectively. (c) $T$ derivative of $\rho$ ($T$) for the SRO thin films under a high magnetic field ($H\perp = 5$ T). (d) Critical exponents $n$ as a function of $T$ under a high magnetic field ($H\perp = 5$ T).

**TABLE I.** Fitting parameters and characteristic $T$s.

|  | $(100)_{tetragonal}$ | $(110)_{monoclinic}$ | $(111)_{trigonal}$ |
|---|---|---|---|
| $T_c$ (K) from $M(T)$ | 147 ± 1.5 | 155 ± 1.5 | 157 ± 1.5 |
| $T_c$ (K) from $\rho(T)$ | 144 ± 1.5 | 154 ± 1.5 | 155 ± 1.5 |
| $A$ | $2.14 \times 10^{-1} \pm 0.01$ | $2.44 \times 10^{-1} \pm 0.01$ | $2.88 \times 10^{-1} \pm 0.01$ |
| $\gamma$ | 0.311 ± 0.01 | 0.287 ± 0.01 | 0.259 ± 0.01 |
| $B$ | $3.31 \times 10^{-4} \pm 9 \times 10^{-6}$ | $1.77 \times 10^{-4} \pm 9 \times 10^{-6}$ | $1.49 \times 10^{-4} \pm 9 \times 10^{-6}$ |
| $J$ ($k_B$ K) | 15.8 ± 2 | 24.0 ± 2 | 26.9 ± 2 |
| $D$ (meVÅ²) @ 0 K | 321.2 ± 6 | 487.9 ± 6 | 546.9 ± 6 |
| $T_m$ (K) | 124 ± 3.5 | 80 ± 3.5 | 75 ± 3.5 |
| $T^*$ | 20.4 ± 3 | 21.9 ± 3 | 23.9 ± 3 |
| @ $H\perp$ = 5 T | | | |
| $T_c$ (K) from $M(T)$ | 186.5 ± 1.5 | 189 ± 1.5 | 190 ± 1.5 |
| $T_c$ (K) from $\rho(T)$ | 180 ± 1.5 | 187 ± 1.5 | 189 ± 1.5 |
| $T_m$ (K) | 35.5 ± 3.5 | 37.2 ± 3.5 | 37.2 ± 3.5 |
| $T^*$ | 32.6 ± 3 | 32.6 ± 3 | 33.7 ± 3 |

# Crystalline symmetry-dependent magnon formation in itinerant ferromagnet SrRuO$_3$

H. I. Seo[*], S. Woo[*], S. G. Jeong, Prof. W. S. Choi

Department of Physics, Sungkyunkwan University, Suwon 16419, Republic of Korea

e-mail : choiws@skku.edu

J. Kim[*], Prof. T. Park

Department of Physics, Sungkyunkwan University, Suwon 16419, Republic of Korea

Center for Quantum Materials and Superconductivity, Department of Physics, Sungkyunkwan University, Suwon 16419, Republic of Korea

[*]These authors contributed equally to this paper.

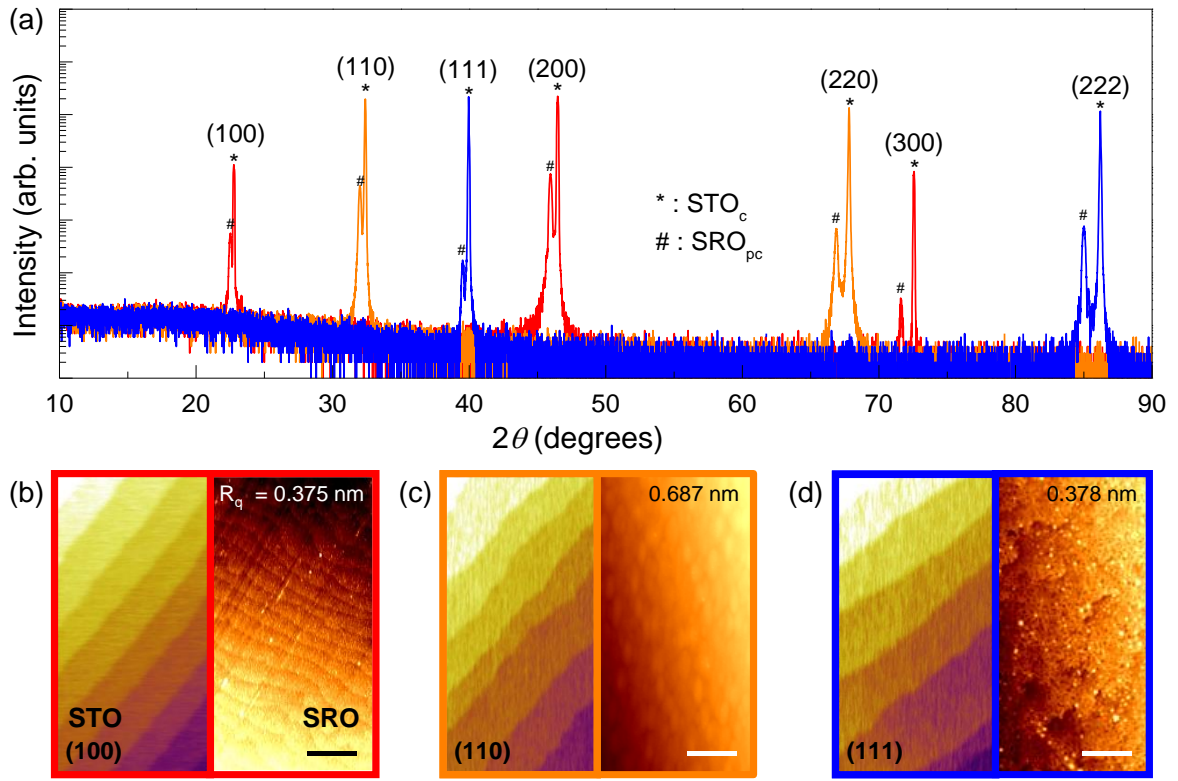

**FIG. S1.** (a) Wider range XRD $\theta$–$2\theta$ scans indicating high-quality single crystalline SRO epitaxial thin films. (b)-(d) Atomic Force Microscopy images of the STO substrates (left) and SRO thin films (right) with (b) (100)-, (c) (110)-, and (d) (111)-surface orientations. All the films show smooth surfaces. The scale bars denote 1 μm.

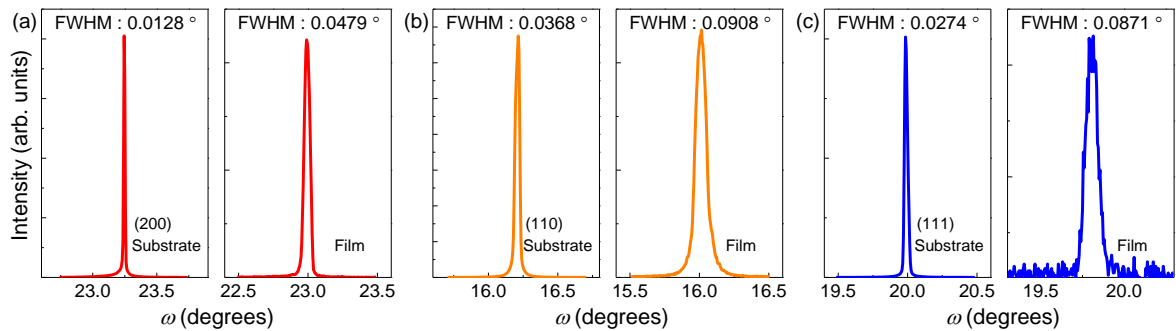

**FIG. S2.** The rocking curve scans of STO substrate (left) and SRO films (right). The FWHM ratio of the film to the substrate are 3.74, 2.47, and 3.18, respectively, for the (a) (100)-, (b) (110)-, and (c) (111)-oriented thin films.

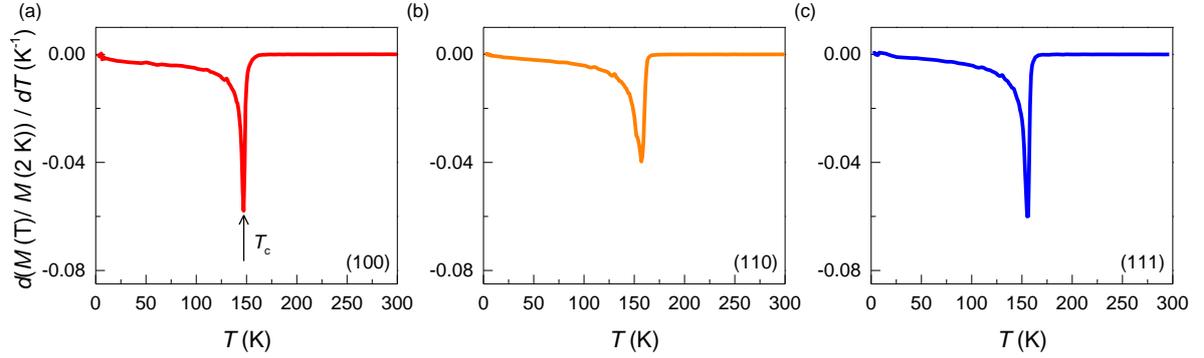

**FIG. S3.** Determination of $T_c$ for (a) (100)-, (b) (110)-, and (c) (111)-oriented SRO thin films, respectively, by using $(1/M) \times (dM/dT)$.

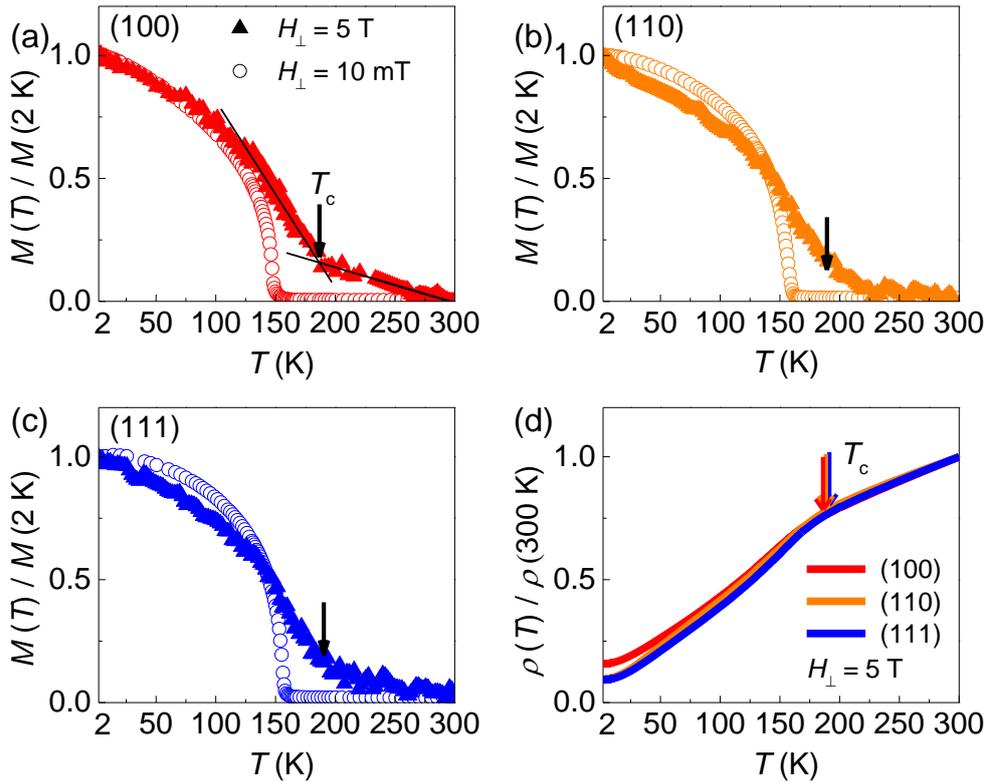

**FIG. S4.** Magnetic and transport behaviors of the SRO thin films under a magnetic field of $H\perp$ = 5 T. Normalized $M(T)$ of (a) (100)-, (b) (110)-, and (c) (111)-oriented SRO thin films. Filled triangles (open circles) denote $M$ at $H\perp$ = 5 T (100 Oe). The $T_c$s are defined as the intersections of the two interpolation lines. (d) $\rho(T)/\rho_{300\,K}$ of the SRO thin films under $H\perp$ = 5 T. The arrows indicate $T_c$.

Appendix

Estimation of the effect of spin dimensionality change to Bloch's law is not straightforward, as Bloch's law is based on Heisenberg model. Because the magnon gap ($\Delta$) in SRO is closely related to the spin dimensionality, we claim that $\Delta$ can be modulated as the consequence of spin dimensionality change, which could modify the Bloch's law. We additionally assume that the change in the magnon DOS is small. Typical Bloch' law can be written as,

$$<n> = \frac{1}{e^{\beta \hbar \omega} - 1}, \text{ where } \beta = \frac{1}{k_B T} \text{ and } \Sigma n = \int_0^\infty d\omega \, D(\omega) <n>.$$

The total number of magnon is given by,

$$\Sigma n = \int_0^\infty d\omega \, D(\omega) <n> = \frac{1}{4\pi^2} \left(\frac{\hbar}{2JSa^2}\right)^{3/2} \int_0^\infty \frac{\omega^{1/2}}{e^{\beta(\hbar\omega)} - 1} d\omega.$$

The integral term is finite and gives the value $4\pi^2(0.0587)$ and $(1/\beta)^{3/2}$ term emerges, leading to the $T^{3/2}$-behavior.

By introducing $\Delta$, one can rewrite the magnon distribution as,

$$<n> = \frac{1}{e^{\beta(\hbar\omega - \Delta)} - 1},$$

$$\Sigma n = \frac{1}{4\pi^2}\left(\frac{\hbar}{2JSa^2}\right)^{\frac{3}{2}} \left(\frac{1}{\beta \hbar}\right)^{\frac{3}{2}} \int_\Delta^\infty \frac{x^{1/2}}{e^{-\beta\Delta}e^x - 1} dx,$$

where $x = \beta\hbar\omega$. The essential $T^{3/2}$-behavior is unchanged and only the integral term changes slightly to $4\pi^2(0.0649)$, by assuming $\Delta = 1$ meV for SRO. This leads to a 10% uncertainty in the $J$ value.